\documentclass[aps,showpacs,preprintnumbers,amsmath,amssymb]{revtex4}
\usepackage{dcolumn}
\usepackage[dvips]{epsfig}
\usepackage{graphicx}
\usepackage{amssymb}
\usepackage{color}
\usepackage{enumerate}

\begin{document}
\baselineskip=0.8 cm
%\preprint{example}
\title{\bf Absorption cross section and Hawking radiation of the electromagnetic field with Weyl corrections}
%\altaffiliation{}
\author{Hao Liao, Songbai Chen\footnote{csb3752@hunnu.edu.cn}, Jiliang Jing
\footnote{jljing@hunnu.edu.cn}}
\affiliation{College of Physics and Information Science, Hunan
Normal University, Changsha, Hunan 410081, P. R. China\\
Key Laboratory of Low Dimensional Quantum Structures \\ and Quantum Control of Ministry of Education,
Hunan Normal University, Changsha, Hunan 410081, P. R. Chins}

\begin{abstract}
\baselineskip=0.6 cm
\begin{center}
{\bf Abstract}
\end{center}

We have investigated the absorption cross section and the Hawking radiation of electromagnetic field with Weyl correction in the background of a four-dimensional Schwarzschild black hole spacetime. Our results show that
the properties of the absorption cross section and the Hawking radiation depend not only on the Weyl correction parameter, but also on the parity of the electromagnetic field, which is quite different from those of the usual electromagnetic field without Weyl correction in the four-dimensional spacetime. With increase of Weyl correction parameter, the absorption probability, the absorption cross section, the power emission spectra and the luminosity of Hawking radiation decreases with Weyl correction parameter for the odd-parity electromagnetic field and increases with the event-parity electromagnetic field.

\end{abstract}

 \maketitle
\newpage

Since generalized Einstein-Maxwell theories contain higher
derivative interactions and carry more information about the electromagnetic field, a lot of attention have been recently focused on studying such kind of generalized Einstein-Maxwell theories in order to probe the full properties and effects of the electromagnetic fields. There are two main classes of the generalized Einstein-Maxwell theories. The first class is minimally coupled gravitational-electromagnetism in which in the Lagrangian there is no coupling
 between the Maxwell part and the curvature part, but the form of Lagrangian of electromagnetic field is changed. The well-known Born-Infeld theory \cite{Born} belongs to such class of generalized Einstein-Maxwell theory. Born-Infeld theory removes the divergence of the electron's self-energy in the classical Maxwell electrodynamics and possesses good physical properties including the absence of shock waves and birefringence phenomena
\cite{Boillat}. Moreover, Born-Infeld theory has also received special
attention because it enjoys an electric-magnetic duality \cite{Gibbons} and can describe gauge fields in the low-energy regime of string
and D-Brane physics \cite{Fradkin}.
In the second class of generalized Einstein-Maxwell theory, there exist the nonminimal coupling terms between the gravitational and electromagnetic fields in the Lagrangian \cite{Balakin,Faraoni,Hehl}.  These nonminimal coupling terms modify the coefficients of the second-order derivatives in the Maxwell and Einstein equations and change behavior of gravitational and electromagnetic waves in the spacetime, which may result in time delays in the arrival of those waves \cite{Balakin}. Moreover, it is also find that such a kind of coupled terms may affect evolution of the early Universe because they may modify electromagnetic quantum fluctuations, which could affect the inflation \cite{Turner,Mazzitelli,Lambiase,Raya,Campanelli}. Furthermore, these cross-terms can be used as attempt to explain the large scale magnetic fields observed in clusters of galaxies \cite{Bamba,Kim,Clarke}.

The theory of electromagnetic field with Weyl corrections contains a coupling between the Maxwell field and the Weyl tensor \cite{Weyl1,Drummond}. Since Weyl tensor is actually related to the curvature tensors $R_{\mu\nu\rho\sigma}$, $R_{\mu\nu}$ and the Ricci scalar $R$, the theory of electromagnetic field with Weyl corrections can be treated as a special kind of generalized Einstein-Maxwell theory with the coupling between the gravitational and electromagnetic fields.
These special coupling terms could be obtained from a calculation in QED of the photon effective action from one-loop vacuum polarization on a curved background \cite{Drummond}. Moreover, it was found that these couplings could exist near classical compact astrophysical objects with high mass density and strong gravitational field such as the supermassive black holes at the center of galaxies \cite{Dereli1,Solanki}. Considering that black hole is an important subject in the modern physics, a lot of efforts have been recently focused on probing the effects of Weyl correction on black hole physics. In Ref.\cite{Weyl1}, the authors studied the effects of Weyl correction on holographic conductivity and charge diffusion in the anti-de Sitter
spacetime and found that the presence of Wely correction changes the universal
relation with the $U(1)$ central charge observed at leading order.
Moreover, the dependence of the holographic superconductors on Weyl corrections are
also studied in \cite{Wu2011,Ma2011,Momeni}. It is shown that
Weyl corrections modify the critical temperature at which holographic superconductors occur \cite{Wu2011} and in the St\"{u}ckelberg mechanism \cite{Ma2011} Weyl corrections also change the order of the phase
transition of the holographic superconductor. In the AdS soliton background, it is find that Weyl corrections also affect the phase transition between the holographic insulator and superconductor \cite{zhao2013}. Recently, we \cite{sb2013} studied the dynamical evolution of the
electromagnetic field with Weyl corrections in the
Schwarzschild black hole spacetime and analyze the effect of the Weyl
corrections on the stability of the black hole. In this letter we are going to study the Hawking radiation of electromagnetic field with Weyl corrections in the background of a Schwarzschild black hole. We will calculate the absorption probability, absorption cross section and the luminosity of Hawking radiation and show physics brought by the Weyl corrections and the parity of electromagnetic field.

Let us now first review briefly the wave equations of the electromagnetic field with Weyl corrections in the background of a black hole \cite{sb2013}. For the electromagnetic field with Weyl corrections, the action in the black hole spacetime can be modified as
\begin{eqnarray}\label{acts}
S=\int d^4x \sqrt{-g}\bigg[\frac{R}{16\pi
G}-\frac{1}{4}\bigg(F_{\mu\nu}F^{\mu\nu}-4\alpha
C^{\mu\nu\rho\sigma}F_{\mu\nu}F_{\rho\sigma}\bigg)\bigg].
\end{eqnarray}
where $F_{\mu\nu}$ is
the usual electromagnetic tensor, which is related to the electromagnetic
vector potential $A_{\mu}$ by $F_{\mu\nu}=A_{\nu;\mu}-A_{\mu;\nu}$. The coefficient $\alpha$ is a coupling constant with dimensions of length squared and the tensor $C_{\mu\nu\rho\sigma}$ is so-called Weyl tensor, which can be expressed as
\begin{eqnarray}
C_{\mu\nu\rho\sigma}=R_{\mu\nu\rho\sigma}-\frac{2}{n-2}(
g_{\mu[\rho}R_{\sigma]\nu}-g_{\nu[\rho}R_{\sigma]\mu})+\frac{2}{(n-1)(n-2)}R
g_{\mu[\rho}g_{\sigma]\nu},
\end{eqnarray}
where $n$ and $g_{\mu\nu}$ are the dimension and metric of the spacetime, and brackets around indices refers to the antisymmetric part. Obviously, Weyl tensor $C_{\mu\nu\rho\sigma}$ is a function of the Riemann tensor $R_{\mu\nu\rho\sigma}$, the Ricci tensor $R_{\mu\nu}$ and the Ricci scalar $R$. Therefore, the Weyl correction to electromagnetic field can be treated as a kind of special couplings between the gravitational and electromagnetic fields.

Varying the action (\ref{acts}) with respect to $A_{\mu}$, one can find
the corresponding Maxwell equation becomes
\begin{equation}\label{WE}
\nabla_{\mu}(F^{\mu\nu}-4\alpha C^{\mu\nu\rho\sigma}F_{\rho\sigma})=0.
\end{equation}
It is well known that the metric of a Schwarzschild black hole spacetime has a form
\begin{equation}\label{fd}
ds^2=fdt^2-\frac{1}{f}dr^2-r^2d\theta^2-r^2sin^2\theta d\phi^2,
\end{equation}
where $f=1-\frac{2M}{r}$.
In such kind of static and spherical
symmetric black hole background, one can expand $A_{\mu}$ in vector spherical harmonics \cite{rr}
\begin{equation}
A_{\mu}
=\sum_{l,m}
\left( \left[
\begin{matrix}
0 \\
0 \\
\frac{a^{lm}(t,r)}{sin\theta}\partial_{\phi}Y_{lm} \\
-a^{lm}(t,r)sin\theta\partial_{\theta}Y_{lm}
\end{matrix}
\right]
+
\left[
\begin{matrix}
j^{lm}(t,r)Y_{lm} \\
h^{lm}(t,r)Y_{lm} \\
k^{lm}(t,r)\partial_{\theta}Y_{lm} \\
k^{lm}(t,r)\partial_{\phi}Y_{lm}
\end{matrix}
\right]\right),\label{Au}
\end{equation}
where the first term in the right side has parity $(-1)^{l+1}$ and
the second term has parity $(-1)^{l}$, $l$ is the angular quantum
number and $m$ is the azimuthal number. Making use of the following form
\begin{eqnarray}\label{w5}
a^{lm}(t,r)&=&a^{lm}(r)e^{-i\omega t},~~~~~h^{lm}(t,r)=h^{lm}(r)e^{-i\omega t},\nonumber\\
j^{lm}(t,r)&=&j^{lm}(r)e^{-i\omega t},~~~~~k^{lm}(t,r)=k^{lm}(r)e^{-i\omega t},
\end{eqnarray}
and inserting the above expansion (\ref{Au})
into the generalized Maxwell equation (\ref{WE}), we can obtain three independent coupled differential equations. Eliminating $k^{lm}(r)$, we find that equations of motion of electromagnetic field can be decoupled into a single second order differential equation
\begin{eqnarray}
\frac{d^2\Psi(r)}{dr^2_*}+[\omega^2-V(r)]\Psi(r)=0,\label{radial}
\end{eqnarray}
where the tortoise coordinate $r_{*}$ is defined as $dr_*=\frac{
r}{r-2M}dr$. The wavefunction $\Psi(r)$ is a linear combination of the
functions $j^{lm}(r)$, $h^{lm}(r)$, and $a^{lm}(r)$, which appeared
in the expansion (\ref{w5}). Both of the forms of wavefunction $\Psi(r)$ and effective potential $V(r)$ depend on the parity of electromagnetic field. For the odd parity $(-1)^{l+1}$, $\Psi(r)$ and $V(r)$ are given by
\begin{eqnarray}
\Psi(r)_{odd}&=&\sqrt{1-\frac{8\alpha M}{r^3}} ~a^{lm}(r),\label{podd}
\\
V(r)_{odd}&=&(1-\frac{2M}{r})\bigg[\frac{l(l+1)}{r^2}(\frac{r^3+16\alpha
M}{r^3-8\alpha M})-\frac{24\alpha M(2r^4-5Mr^3-10\alpha Mr+28\alpha
M^2)}{r^3 (r^3-8\alpha M)^2}\bigg].\label{vodd}
\end{eqnarray}
For the even parity $(-1)^{l}$, the forms of $\Psi(r)$ and $V(r)$ become
\begin{eqnarray}
\Psi(r)_{even}&=&\frac{r^{\frac{7}{2}}}{l(l+1)}\bigg(-i\omega
h^{lm}(r)-\frac{dj^{lm(r)}}{dr}\bigg)
\frac{\sqrt{r^3+8\alpha M}}{
r^3+16\alpha M},\label{peven}
\\
V(r)_{even}&=&(1-\frac{2M}{r})\bigg[\frac{l(l+1)}{r^2}(\frac{r^3-8\alpha
M}{r^3+16\alpha M})+\frac{24\alpha M(2r^4-5Mr^3+2\alpha Mr+4\alpha
M^2)}{r^3 (r^3-8\alpha M)^2}\bigg].\label{veven}
\end{eqnarray}
It is obvious that the Weyl corrections change the behavior of effective potentials $V(r)_{odd}$ and $V(r)_{even}$. Moreover, the change of effective potential originating from Weyl
corrections is different for the electromagnetic fields with
different parities. For fixed
$l$, we find in Ref.\cite{sb2013} that the peak height of the potential barrier increases with the
coupling constant $\alpha$ for $V(r)_{odd}$ and decreases for
$V(r)_{even}$. It implies that the effects of Weyl
corrections on the absorption cross section and Hawking radiation for the electromagnetic field with the odd parity are different from those of the
field with the even parity. From Eq.(\ref{vodd}), one can find that the effective potential $V(r)_{odd}$ has a discontinuity of the second kind at the point $r_d=(8\alpha M)^{1/3}$ for the positive $\alpha$ and near the  discontinuity point the wave function $\Psi(r)$ is not well-defined. However, if the discontinuity point is located in the region inside the event horizon, the above problem can be avoided in the physical region of the black hole (i.e., $r>r_H$) and then one can study the Hawking radiation of electromagnetic field with Weyl corrections by the standard methods in this case. Therefore, we here must impose a constraint on the value of the coupling constant, $r_H>r_d$ (i.e., $\alpha<M^2$), to keep the continuity of the effective potential and the well-defined behavior of the wave function $\Psi(r)$ in the physical region of black hole for the electromagnetic field with the odd parity. After similar analyses, we can find that the coupling
constant $\alpha$ must be limited in the range $r^3_H-8\alpha M>0$  and $r^3_H+16\alpha M>0$ (i.e., $-\frac{M^2}{2}<\alpha<M^2$) for the field with the even parity.

In order to calculate the absorption cross section and the luminosity of Hawking radiation for the electromagnetic field with Weyl corrections, we
must solve the radial equation (\ref{radial}) numerically. Near
the horizon regime and at infinity, one can find that the radial wavefunction $\Psi(r)$ satisfies the boundary conditions
\begin{eqnarray}\label{bc}
\Psi(r)\approx  \bigg\{
         \begin{array}{rrr}
   A^{tr}_{\omega l}(\omega)e^{-i\omega r_*},\ \ \ \ \ \ \ \ \ \ \ &  & r_*\rightarrow -\infty;\\ \\
     A^{in}_{\omega l}(\omega)e^{-i\omega r_*}+A^{out}_{\omega l}(\omega)e^{i\omega r_*}& , & r_*\rightarrow +\infty,
          \end{array}
\end{eqnarray}
respectively. The coefficients $A^{tr}_{l}(\omega)$, $A^{out}_{l}(\omega)$ and $A^{in}_{l}(\omega)$ obey the conservation relationship
 $|A^{tr}_{l}(\omega)|^2+|A^{out}_{l}(\omega)|^2=|A^{in}_{l}(\omega)|^2$.
With help of this relationship and equation (\ref{bc}), we can calculate the absorption probability
\begin{eqnarray}
A_{l}=\bigg|\frac{A^{tr}_{\omega l}(\omega)}{A^{in}_{\omega l}(\omega)}\bigg|^2=1-\bigg|\frac{A^{out}_{\omega l}(\omega)}{A^{in}_{\omega l}(\omega)}\bigg|^2.
\end{eqnarray}
The absorption cross section $\sigma_{abs}$ is related to the absorption probability $A_{l}$ by
\begin{eqnarray}
\sigma_{abs}=\sum^{\infty}_{l=0}\sigma_{l}
=\frac{\pi}{\omega^2}\sum^{\infty}_{l=0}(2l+1)A_{l}.
\end{eqnarray}
\begin{figure}[htp]
\begin{center}
\includegraphics[width=5.5cm]{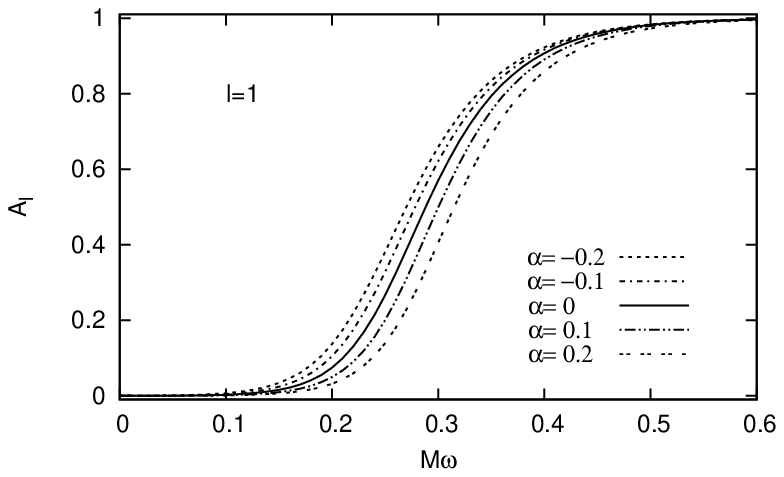}
\includegraphics[width=5.5cm]{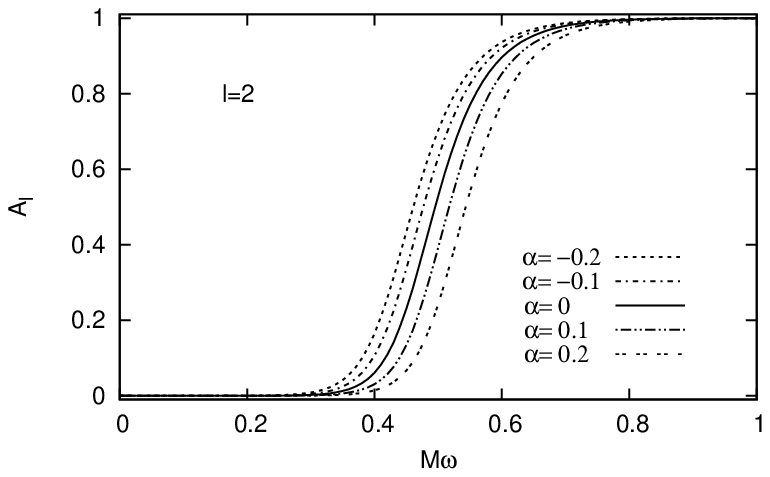}
\includegraphics[width=5.5cm]{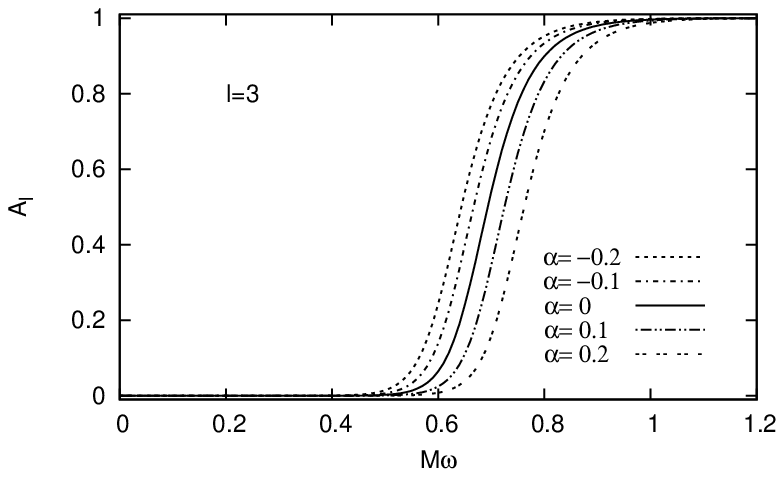}
\caption{Effects of Weyl correction on the absorption probability of electromagnetic field with odd-parity in the Schwarzschild black hole spacetime. Here we set $M=1$.}
\end{center}
\end{figure}
\begin{figure}[htp]
\begin{center}
\includegraphics[width=5.5cm]{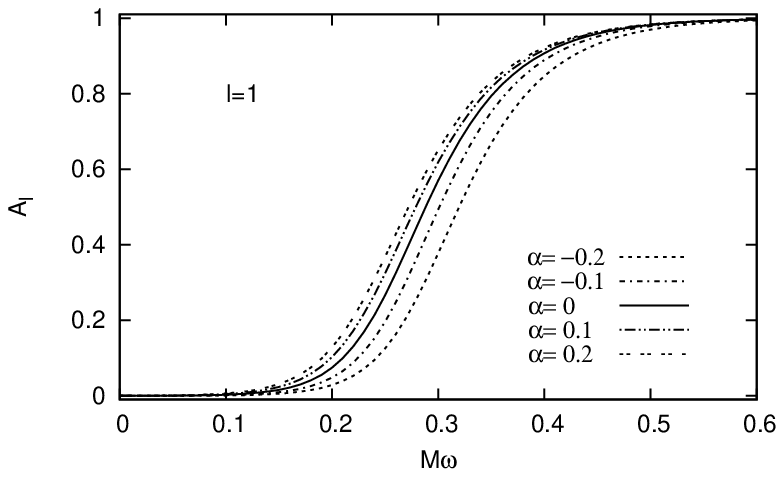}
\includegraphics[width=5.5cm]{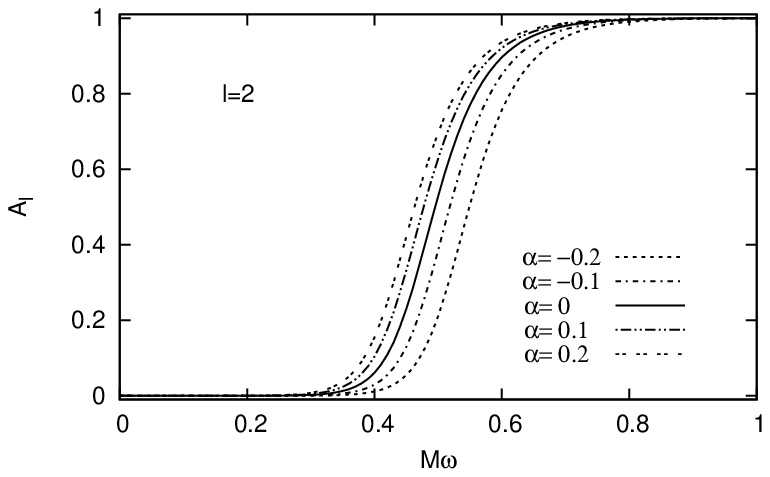}
\includegraphics[width=5.5cm]{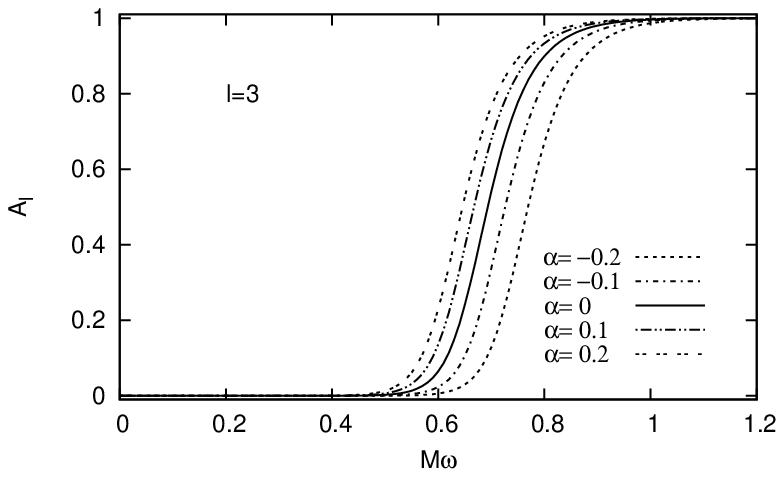}
\caption{Effects of Weyl correction on the absorption probability of electromagnetic field with even-parity in the Schwarzschild black hole spacetime. Here we set $M=1$.}
\end{center}
\end{figure}
In Figs.(1) and (2), we present the effects of  Weyl correction on the absorption probability of electromagnetic field for fixed angular index $l$. For the odd parity electromagnetic field, one can easily see that the absorption probability decreases with the increase of the Weyl correction parameter $\alpha$. The main reason behind this phenomenon is that the
larger $\alpha$ yields the higher peak of the effective potential so
that less electromagnetic wave from infinity can be transmitted to the black hole. For the even parity electromagnetic field, we find that the absorption probability increases with the Weyl correction parameter $\alpha$, which means that the dependence of the absorption probability on the Weyl correction parameter $\alpha$ is different entirely from that of in the case of the electromagnetic field with odd parity. It can be attributed to the difference in the change of the effective potential with the parameter $\alpha$ for the electromagnetic fields with different types of parities. With increase of the angular index $l$, we find that the absorption probability decreases for the two different parities, which is similar to that in the case without Weyl corrections.
\begin{figure}[htp]
\begin{center}
\includegraphics[width=5.5cm]{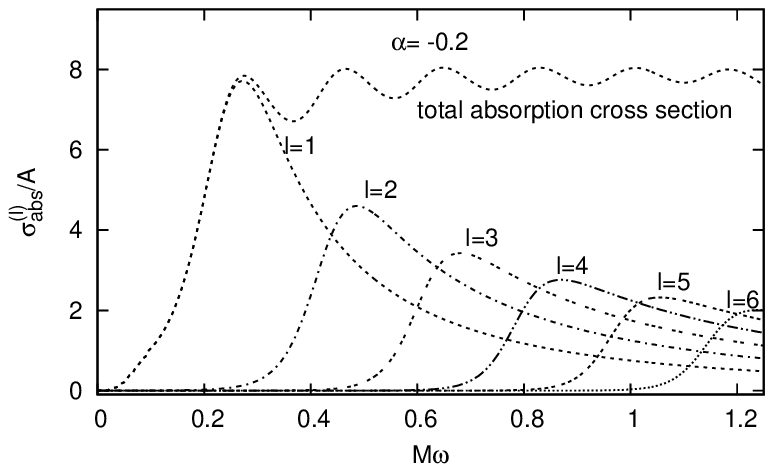}
\includegraphics[width=5.5cm]{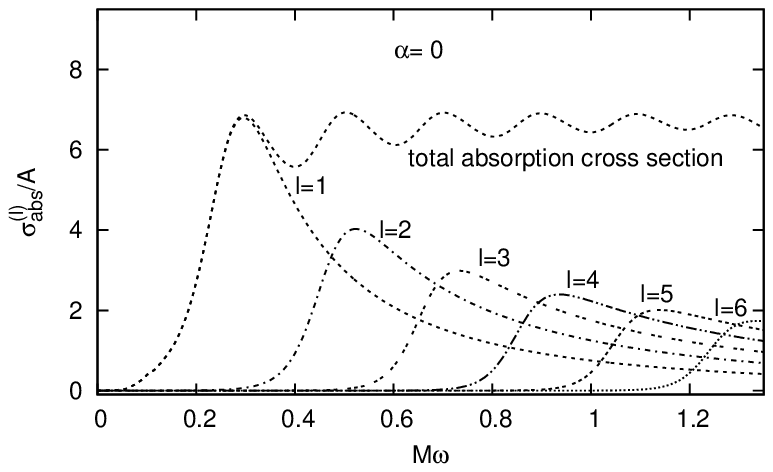}
\includegraphics[width=5.5cm]{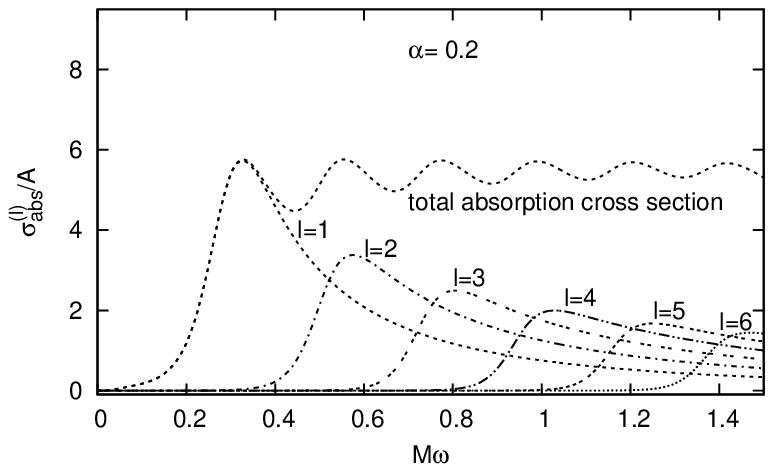}
\caption{Dependence of absorption cross section of the odd-parity electromagnetic field on Weyl corrections in Schwarzschild spacetime. Here we set $M=1$ and $A=4\pi M^2$.}
\end{center}
\end{figure}
\begin{figure}[htp]
\begin{center}
\includegraphics[width=5.5cm]{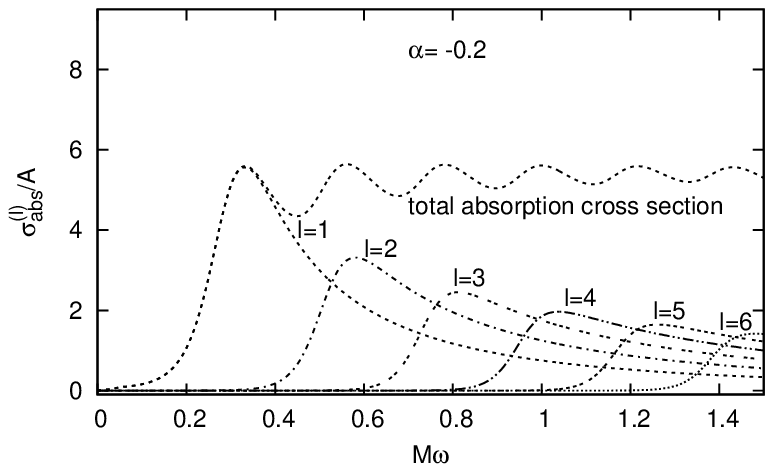}
\includegraphics[width=5.5cm]{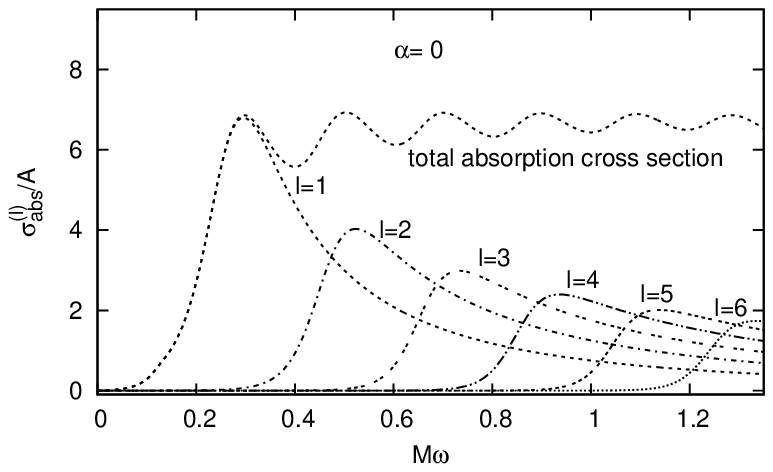}
\includegraphics[width=5.5cm]{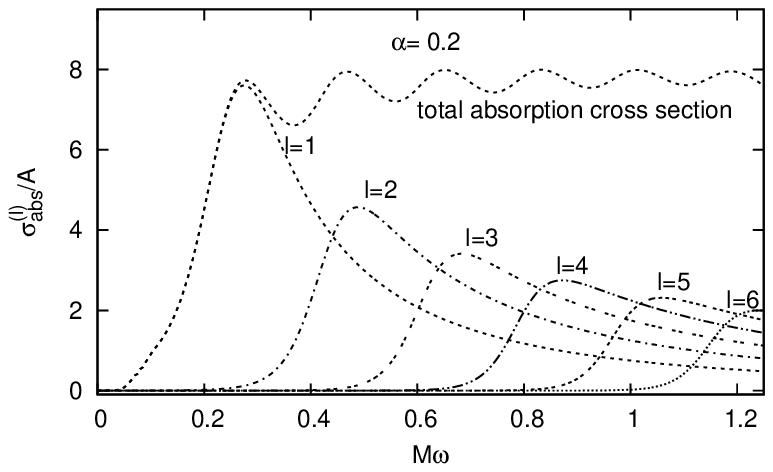}
\caption{Dependence of absorption cross section of the even-parity electromagnetic field on Weyl corrections in Schwarzschild spacetime. Here we set $M=1$ and $A=4\pi M^2$.}
\end{center}
\end{figure}
\begin{figure}[htp]
\begin{center}
\includegraphics[width=6cm]{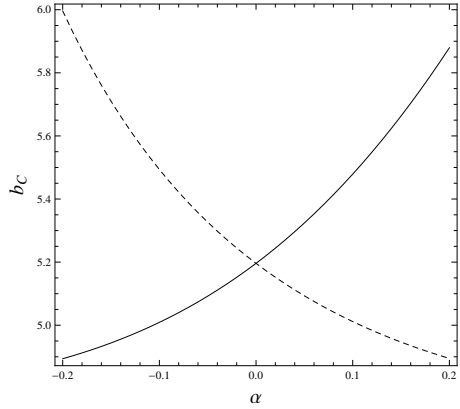}
\caption{The change of $b_{c}$ with Weyl correction parameter $\alpha$ in the Schwarzschild black hole spacetime. The dashed line and the solid line are for the electromagnetic field with odd-parity and even-parity, respectively. Here we set $M=1$.}
\end{center}
\end{figure}

In Figs. (3) and (4), we plot the change of the partial and total absorption cross sections on the Weyl correction parameter $\alpha$ for the odd-parity and even-parity electromagnetic fields, respectively. With increase of the Weyl correction parameter $\alpha$, we find both of the partial and total absorption cross sections decrease for the odd-parity electromagnetic field, but increase for the even-parity one. As the frequency $\omega\rightarrow 0$,  one can easily obtain that the absorption cross section tend to zero for both of different electromagnetic fields, which is the same as that in the case without Weyl corrections. Figs. (3) and (4) also tell us that in the high-energy region the total absorption cross section oscillate around the geometric-optical limit $\sigma_{\text{geo}}$, which is similar to that in the case without Weyl corrections \cite{ns,ad1}. However, in the case with Weyl corrections, we find that the geometric-optical limit $\sigma_{\text{geo}}$ decreases with $\alpha$ for the odd-parity electromagnetic field and increases for the even-parity one. It is not surprising because the geometric-optical limit $\sigma_{geo}$ can be approximated as a function of impact parameter $b_c$, i.e., $\sigma_{geo}\sim\pi b^2_c$ and  $b_c$ is related to the radius of photon sphere $r_{ps}$, which depends on behavior of effective potential of electromagnetic field in the background of black hole spacetime. For the electromagnetic field with Weyl corrections, the change of $b_c$ with $\alpha$ is shown in Fig. (5). It is clear that with increase of the Weyl correction parameter $\alpha$ the impact parameter $b_c$ decreases for the odd-parity electromagnetic field and increases for the even-parity one.

Now let us turn to study the effects of Weyl corrections on Hawking radiation of electromagnetic field in the background of a Schwarzschild black hole.
The Hawking power emission spectrum and the Hawking luminosity of electromagnetic field are \cite{pk,pk1,cmh}
\begin{equation}
\frac{d^2E}{dtd\omega}=\frac{1}{2\pi}\sum_{l}\frac{(2l+1)A_l\omega}
{e^{\omega/T_{H}}-1},
\end{equation}
and
\begin{equation}\label{le1}
L=\sum_{l}\int_{0}^{\infty}\frac{(2l+1)A_l \omega }{e^{\omega/T_{H}}-1}\frac{d\omega}{2\pi},
\end{equation}
respectively,
where $T_H$ is the Hawking temperature of Schwarzschild black hole.
\begin{figure}[htp]
\begin{center}
\includegraphics[width=5.5cm]{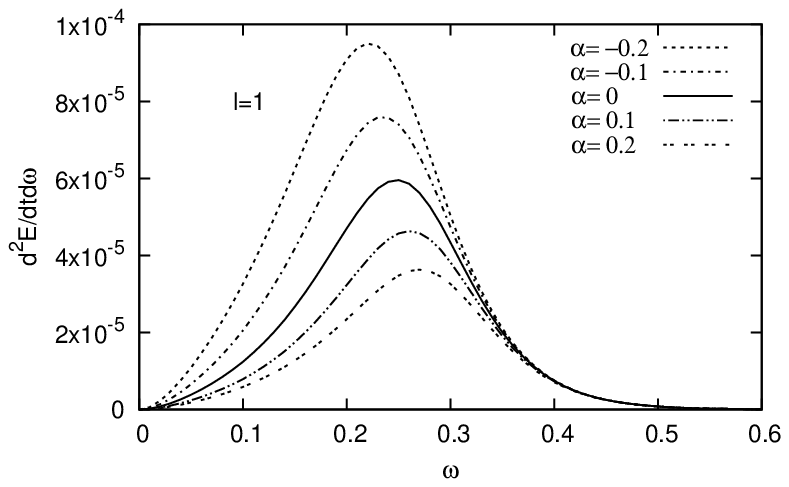}
\includegraphics[width=5.5cm]{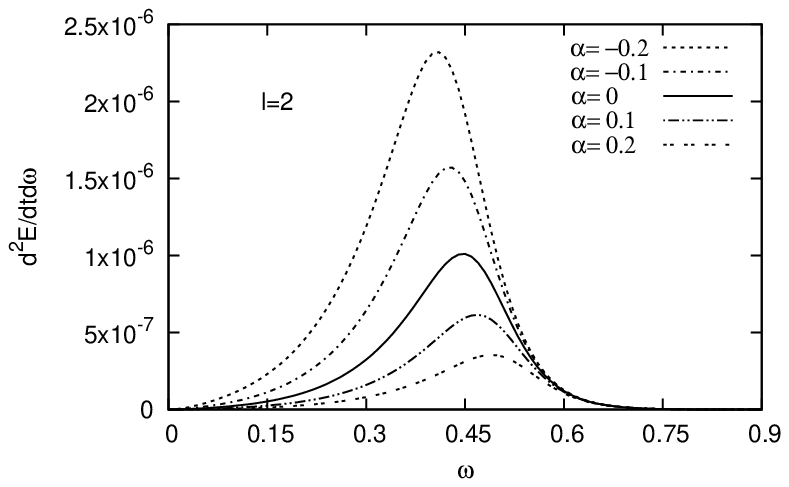}
\includegraphics[width=5.5cm]{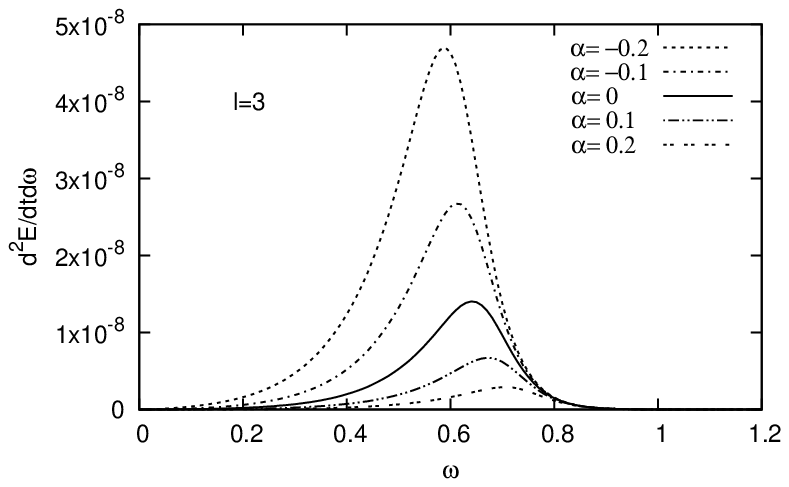}
\caption{Power emission spectra of the odd-parity electromagnetic field with Weyl corrections in the Schwarzschild black hole spacetime. Here we set $M=1$.}
\end{center}
\end{figure}
\begin{figure}[htp]
\begin{center}
\includegraphics[width=5.5cm]{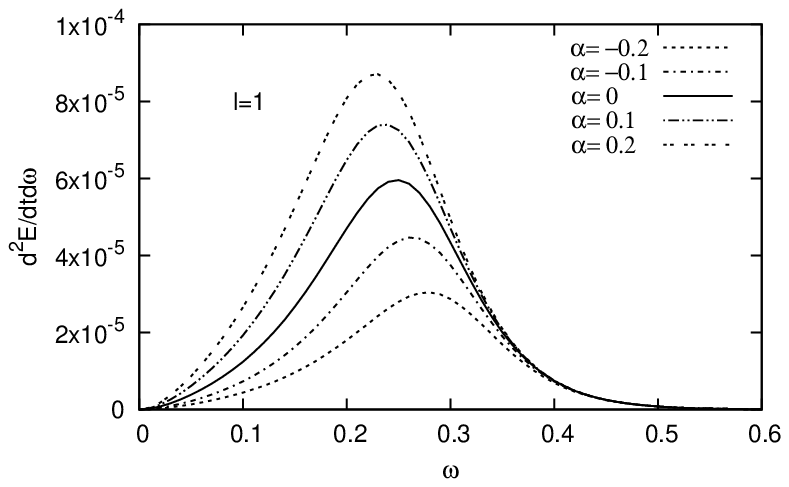}
\includegraphics[width=5.5cm]{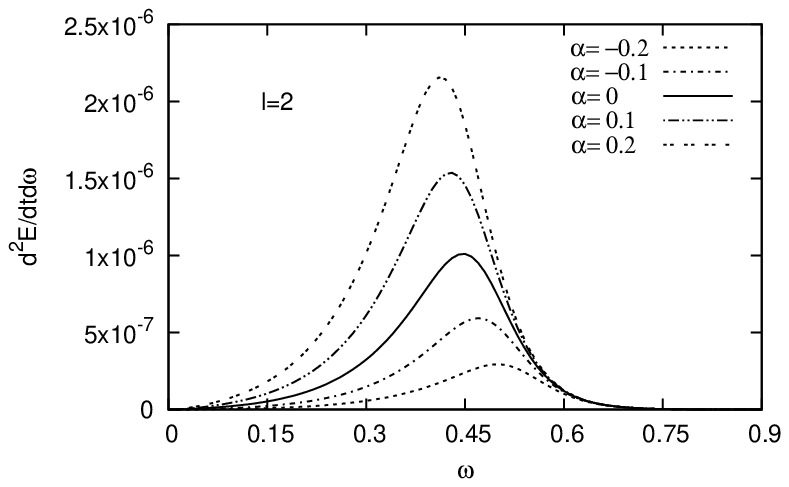}
\includegraphics[width=5.5cm]{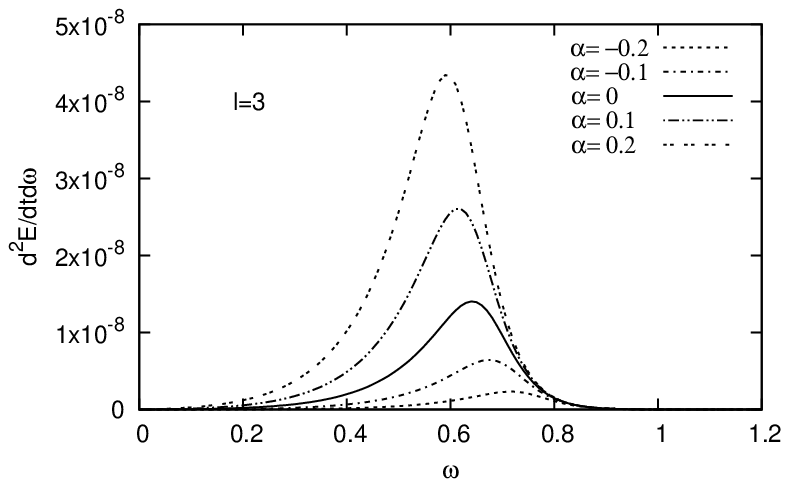}
\caption{Power emission spectra of the even-parity electromagnetic field with Weyl corrections in the Schwarzschild black hole spacetime. Here we set $M=1$.}
\end{center}
\end{figure}
\begin{figure}[htp]
\begin{center}
\includegraphics[width=5.5cm]{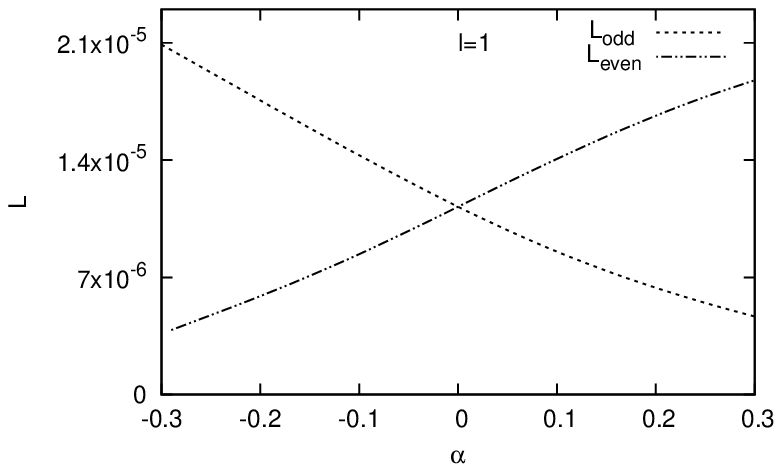}
\includegraphics[width=5.5cm]{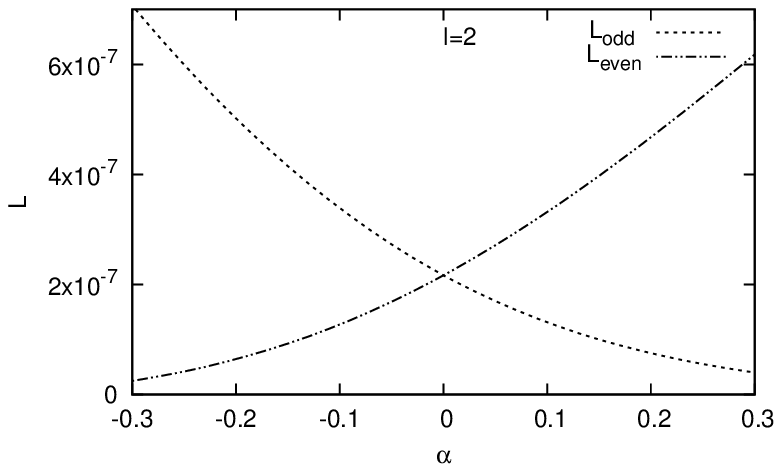}
\includegraphics[width=5.5cm]{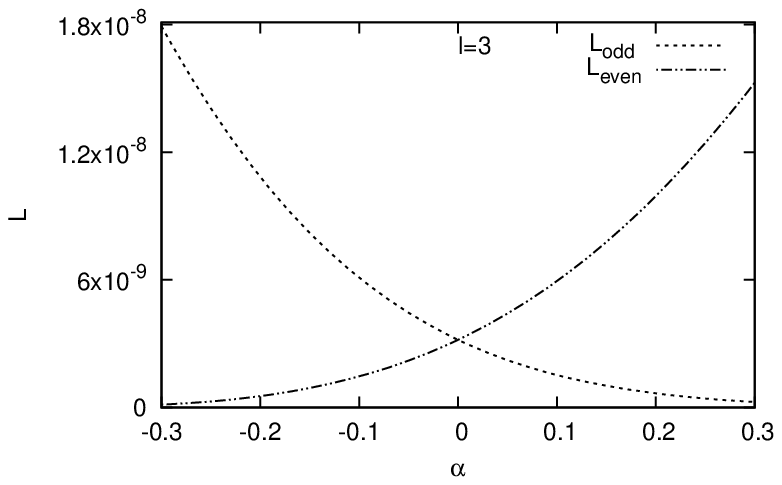}
\caption{Luminosity of Hawking radiation of electromagnetic field with Weyl corrections in Schwarzschild spacetime. Here we set $M=1$.}
\end{center}
\end{figure}
In Figs. (6) and (7), we present the power emission spectra of the electromagnetic field with Weyl corrections in the Schwarzschild black hole spacetime for fixed $l$. With increase of Weyl correction parameter $\alpha$, we find that the power emission spectra decreases for the electromagnetic field with odd parity and increases for the electromagnetic field with even parity. The dependence of power emission spectra on Weyl correction parameter is similar to the dependence of absorption probability on Weyl correction parameter. In Fig. (8), we show the dependence of the luminosity
of Hawking radiation on parameter $\alpha$ for fixed angular index $l$. It is clear that as $\alpha$ increases,  the luminosity $L$ of partial wave decreases for the odd parity electromagnetic field and increases for the even parity one. It is shown again that Weyl corrections modifies the standard results of Hawking radiation of electromagnetic field in the black hole spacetime.

In summary, we have investigated numerically the absorption cross section and the Hawking radiation of an electromagnetic field with Weyl correction in the background of a four-dimensional Schwarzschild black hole spacetime. Our results show that Weyl correction modifies the standard results of the absorption cross section and the Hawking radiation for the electromagnetic field. Due to the presence of Weyl corrections, the properties of the absorption cross section and the Hawking radiation depend not only on the Weyl correction parameter $\alpha$, but also on the parity of the electromagnetic field. With increase of Weyl correction parameter $\alpha$, we find that both of the absorption probability and the absorption cross section, decreases with Weyl correction parameter for the odd-parity electromagnetic field and increases with the event-parity electromagnetic field. In the low frequency limit $\omega\rightarrow 0$, we find that both of the absorption probability and the absorption cross section tend to zero, which is similar to those in the case of electromagnetic field without Weyl correction. In high-energy region we also find that the total absorption cross section oscillates around the geometric-optical limit $\sigma_{\text{geo}}$. However, in the case with Weyl corrections, the geometric-optical limit $\sigma_{\text{geo}}$ also depend on the Weyl correction parameter $\alpha$ and the parity of the electromagnetic field. Moreover, we also find that the power emission spectra and the luminosity of Hawking radiation decreases with Weyl correction parameter for the odd-parity electromagnetic field and increases with the event-parity electromagnetic field. Our results show again that Weyl corrections modifies the properties of the absorption cross section and the Hawking radiation for the electromagnetic field in the black hole spacetime.

{\bf Acknowledgments}

This work was  partially supported by the National Natural Science Foundation of China under Grant No.11275065, the NCET under Grant
No.10-0165, the PCSIRT under Grant No. IRT0964,  the Hunan Provincial Natural Science Foundation of China (11JJ7001) and the construct
program of key disciplines in Hunan Province. J. Jing's work was
partially supported by the National Natural Science Foundation of
China under Grant Nos. 11175065, 10935013; 973 Program Grant No.
2010CB833004.

\vspace*{0.2cm}

%\newpage

\end{document}